# Breaking the Boundaries of Knowledge Space: Analyzing the Knowledge Spanning on the Q&A Website through Word Embeddings

**Haochuan Cui [a], Tiewei Li [b,*], Cheng-Jun Wang [c,*]**

[a] School of Systems Science, Beijing Normal University, Beijing, China,

haochuancui@mail.bnu.edu.cn

[b] School of Journalism and Communication, Nanjing University, Nanjing, China,

litiewei219@163.com

[c] School of Journalism and Communication, Nanjing University, Nanjing, China,

wangchengjun@nju.edu.cn

[*] Corresponding author. E-mail addresses: litiewei219@163.com (T. Li) &

wangchengjun@nju.edu.cn (C.J. Wang)

## Abstract

The challenge of raising a creative question exists in recombining different categories of knowledge. However, the impact of recombination remains controversial. Drawing on the theories of knowledge recombination and category spanning, we propose that both the distance of knowledge spanning and the hierarchy of knowledge shape the appeal of questions. Using word embedding models and the data collected from a large online knowledge market (N = 463,545), we find that the impact of knowledge spanning on the appeal of questions is parabolic: the appeal of questions increases up to a threshold, after which point the positive effect reverses. However, the nonlinear influence of knowledge spanning is contingent upon the hierarchy of knowledge. The theoretical and practical implications of these findings for future research on knowledge recombination are discussed. We fill the research gap by conceptualizing question asking as knowledge spanning and highlighting the theoretical underpinnings of the knowledge hierarchy.

## 1. Introduction

Knowledge begins with asking questions (Gazan, 2010; Ravi et al., 2014; Shi et al., 2021). Knowledge is one kind of public goods. To avoid the tragedy of the commons (Hardin, 1968; Ostrom, 1990) and enable, support, and facilitate the sharing of knowledge (Stewart, 1996), the knowledge market is established. The question-and-answer (i.e., Q&A) website is one kind of knowledge market. According to the types of payments, the Q&A websites can be classified into two categories: fee-based knowledge markets (e.g., Google Answers) and free knowledge markets (e.g., Quora, Zhihu). The free knowledge exchange model works by increasing the reputation of answers and questioners. How to ask creative questions plays an essential role in attracting high-quality answers. The theories of knowledge recombination suggest that boundary spanning plays a crucial role in asking novel questions (Schumpeter, 1934; Nelson, et al., 1982; Moaniba, et al., 2018; Guan, et al., 2018; Zhang, et al., 2019).

However, there are a number of research gaps in prior research on knowledge spanning. First, there is a debate on the role played by knowledge spanning (Nelson & Winter, 1982; Kuhn, 1977; Ferguson & Carnabuci, 2017). On the one hand, knowledge spanning can bring a broader niche market and help cultural products increase the potential for extraordinary achievements (Keuschnigg & Wimmer, 2017; Hsu, et al., 2012; Ordanini, et al., 2018). On the other hand, spanning multiple categories is both challenging and risky (Ferguson & Carnabuci, 2017; Hsu, et al., 2009). Therefore, it faces an even more considerable risk of failure. Second, prior research primarily focuses on knowledge recombination in scientific or technological innovations. There is a lack of research about the knowledge recombination existing in the daily life of ordinary people (Badilescu-Buga, E., 2013; Shin, et al., 2001; Shi et al., 2021). Third, although prior research highlights the significance of capturing the extent of recombination in the knowledge space (Moaniba,et al., 2018), there is a lack of methods to explicitly model the knowledge space (Morita, et al., 2004; Min, et al., 2021). Fortunately, based on neural networks and deep learning, word embedding models provide a powerful lens to represent the knowledge space and measure the distance of knowledge spanning (Mikolov et al., 2013a; Mikolov et al., 2013b; Tang, et al., 2019). Fourth, most studies focus on the similarity of category spanning, the hierarchy of knowledge is largely ignored (Cole, 1983; Hayakawa & Hayakawa, 1939; Lerner & Lomi, 2018; Tang, et al., 2019). To fill these research gaps, this study aims to solve the following puzzlement: what are the roles played by knowledge spanning and knowledge hierarchy in asking attractive questions in the online knowledge market?

Drawing on the theories of recombination (Schumpeter, 1934; Nelson & Winter, 1982) and category spanning (Hannan & Freeman, 1977; Hannan et al., 2007; Hannan, 2010; Hsu et al., 2009), we claim that both the distance of knowledge spanning and the hierarchy of knowledge shape the appeal of questions on the Q&A website. We collect a novel dataset from a large Q&A website and address these challenges as mentioned above by quantifying the distance of knowledge spanning and hierarchy of knowledge using the word embedding methods (Mikolov et al., 2013a; Mikolov et al., 2013b; Kozlowski et al., 2019; Li, et al. 2020; Lee, et al., 2021). The findings support the existence of a parabolic relationship between knowledge spanning and the appeal of questions. More interestingly, the knowledge hierarchy moderates the nonlinear impact of knowledge spanning. In the following sections, we will first outline the theories of knowledge recombination, reformulate our research puzzlement, and propose the operational hypotheses. Second, we introduce the research methods and the findings of this present study. Third, we summarize the findings and discuss the implications.

## 2. Theoretical Framework

### 2.1 Theories of Knowledge Recombination

Creativity is stimulated when diverse ideas are mingled (Schumpeter, 1934; Nelson & Winter, 1982). According to the seminal work of Nelson and Winter (1982), "The creation of any sort of novelty in art, science, or practical life–consists to a substantial extent of a *recombination* of conceptual and physical materials that were previously in existence" (Nelson and Winter,1982, p.130). Similarly, Koestler coined the term "bisociation" to elucidate the logic of recombination and the nature of creativity (Koestler, 1964). Koestler (1964) asserts that human beings are used to thinking within a single frame of thought. Nevertheless, creative ideas lie in the intersection area of two frames of thought. The act of creation operates through blending the ingredients from two seemingly incompatible frames of thought. Based on the concept of bisociation, Koestler (1964) examines the creative activity in humor, art, and scientific research. In all, innovation often means reassessing and rearranging the existing knowledge, techniques, concepts, and materials. Therefore, the freedom to explore new directions or divergent thinking is necessary.

However, divergent thinking is not enough. Kuhn (1977) argues that both convergent thinking and divergent thinking are essential to innovation. The essential tension between these two modes of thought generates the best scientific research (Kuhn, 1977). Most problems can be answered using the existing theories and methods. Under normal conditions, people are not innovators but problem solvers. "Only investigations firmly rooted in the contemporary scientific tradition are likely to break that tradition and give rise to a new one" (Kuhn, 1977, p. 227). For example, the essential tension between productive tradition and risky innovation affects a researcher's choice of research problems. Ultimately, the work bound by tradition invariably changes the tradition (Kuhn, 1977). In other words, It is easier to break the tradition from the inside. Kuhn asserts that the most fundamental advance in science depends on recognizing the causes of the crisis. The convergent tools and standards developed by well-defined tradition help people better identify the loci of troubles. The collision between theoretical expectations and unexpected observations brings breakthrough discoveries (Peirce, 2016).

In general, the factors affecting knowledge recombination (e.g., how people ask questions) can be classified into at least three categories. At the individual level, it is related to a person's interest, taste, education (e.g., training), career trajectory, new resources (e.g., mentor, collaborators, expertise, information), and the expenditure of effort. At the local level, it is influenced by institutional factors and disciplinary culture. At the social level, it is also influenced by the social structure and policies. According to Foster et al. (2015), all these factors can be well considered within Bourdieu's field theory of science (Bourdieu, 1975, 2004; Foster et al., 2015).

Bourdieu's field theory of science provides an organizing framework to understand knowledge recombination. According to Bourdieu, agents' habitus and capital in the field shape their practice and lifestyles (Bourdieu, 1975). The habitus is an acquired system of habits, tastes, dispositions, skills, and expectations. It determines how individuals perceive and react to the social world. As the structured structure (cognitive schemes), the habitus is the principle organizing the perception of the social world. As the structuring structure (generating schemes), habitus organizes practices and perceptions. The interplay between agents' positions and their habitus guides their actions. Taking scientific research as an example, researchers strive to accumulate scientific capital and occupy the dominant positions in specific fields (Guan et al., 2017; Guan et al., 2018; Zhang et al., 2019). The strategic choice of research question continually re-creates tradition and punishes deviance in the field (Bourdieu, 1975). When knowledge has been crystallized into stable traditions, a disposition towards the tradition helps maximize productivity and avoid unnecessary risks.

Knowledge recombination can be viewed as a reshuffling process of the knowledge network. Uzzi and Spiro (2015) argue that the small-world network affects human behaviors by influencing connectivity and cohesion. In a small-world network, people interact within local clusters, and long-ranged ties connect these local clusters. These properties of the small-world network promote the diffusion of ideas from one cluster to another in the social system. The small-world network is one kind of recombination. It can be realized by rewiring the links in a regular network with a probability (Watts & Strogatz, 1998). For the regular network, a node only connects with its nearby neighbors. In this case, there is little room for innovation. By randomly rewiring the links, knowledge recombination is realized by adding long-ranged ties while preserving local clustering. However, if the probability of reshuffling is larger than a threshold, the local clustering would drop dramatically, which will hinder knowledge recombination. Thus, there is a tradeoff in choosing the optimal degree of recombination.

Together, there exists a vibrant body of theoretical frameworks to understand knowledge recombination and its outcome. It is reasonable to conclude that knowledge recombination is essential for raising attractive or popular questions. However, as our review of the theories of knowledge recombination suggests, there is a debate on the concrete mechanism and outcome of knowledge recombination. The limitation of knowledge recombination is also apparent. On the one hand, knowledge recombination is risky for the agents to achieve a high position and gain symbolic power. On the other hand, the tradition is productive for supplying a series of theories and methods, identifying the loci of trouble, maximizing productivity, and accumulating symbolic capital (e.g., reputation or peer recognition). Nevertheless, there is no doubt that knowledge recombination plays a critical role in innovation. The key to the problem is the mechanism of knowledge recombination. Thus, the *research puzzle* of knowledge recombination can be reformulated: *what kind of knowledge recombination can bring better communication outcomes?*

## 2.2 Knowledge Sharing on the Q&A Website

The Q&A website constitutes an informal social field of knowledge sharing for the public. With the accelerated updating and iteration of knowledge, individuals and organizations cannot deal with their problem relying on their own knowledge stock (Ostrom, 1990; Gazan, 2010; Qi, et al., 2020). Although scientists and elites can also take part in the Q&A process, most participants are ordinary people. Just as C. Wright Mills has put in his seminal work, "*what ordinary men are directly aware of and what they try to do are bounded by the private orbits in which they live; their visions and their powers are limited to the close-up scenes of job, family, neighborhoods; in other milieux, they move vicariously and remain spectators*" (Mill, 1959, pp. 3). Digital media has transformed the way to acquire knowledge. The users of the knowledge market can not only share knowledge but also integrate existing knowledge and develop new knowledge through online interaction (Deng, et al., 2020; Qi, et al., 2020). Everyone can act as the creator, disseminator, and consumer of knowledge. With the aid of the knowledge market (e.g., Q&A website), ordinary people's individual questions are aggregated to a global knowledge space (Deng, et al., 2020). As a result, the hidden individual knowledge becomes visible social knowledge.

In contrast with scientific research and technical innovations, the Q&A website is characterized by creating and spreading social knowledge. For example, when people ask a question on the Q&A website, they may not expect to get a technical answer (Liu, Z., & Jansen, B. J., 2017; Liu, et al., 2018). Just as Charles Tilly has illustrated, technical explanations are only one of four types of reasons adopted by people in daily life (Tilly, 2006). In addition to technical explanations, conventions, stories, and codes (e.g., legal judgments) are three other types of reasons. According to Tilly (2006), when people ask and answer questions, they confirm, repair, claim, or deny the social relations among them.

Therefore, social factors also matter for knowledge sharing on the Q&A website (Fu et al., 2019). The social factors extensively investigated in prior research can be classified into three categories:

First, the attributes of users shape knowledge sharing. Users' online profile, posting behavior, language style, and social activities reflect their intention of knowledge sharing. Liu et al. (2017) find that these non-Q&A features can achieve a 70% success rate. More importantly, prior studies suggest that authors' positions in the collaboration network has a significant influence on a paper's citations (Abbasi et al., 2011; Li et al., 2013). Further, Guan et al. (2018) show that researchers' structural holes in both collaboration networks and knowledge networks have a positive impact on knowledge creation. Similarly, Colladon et al. (2020) show that the future success of scientific research can be predicted through social network and semantic analysis. Moreover, the reputation of contributors has a reversed U-shaped influence on the popularity of answers (Osatuyi et al., 2022).

Second, the content features (e.g., level of details, specificity, and clarity) can promote knowledge sharing. For example, Chua et al. (2015) analyze the metadata and content of the questions collected from Stack Overflow. They find the level of details, specificity, clarity, and the socio-emotional value of questions are important content features for predicting whether a question would be answered or not (Chua et al., 2015). Further, the content quality and source credibility is useful for finding good answers in online knowledge communities (Neshati, 2017). Osatuyi et al. (2022) find that the information quality of answers has a positive effect on their probability of being selected as the best answer. However, the returns of information quality diminish quickly. Sussman et al. (2003) extend the elaboration likelihood model to computer-mediated communication and propose the knowledge adoption model (KAM). They distinguish the central cues (e.g., content quality) from peripheral cues (source credibility) and posit that information cues' influence on information adoption is mediated by perceived information usefulness (Sussman et al., 2003). Liu X. et al. (2020) propose a text analytic framework based on related theories (Petty et al., 1986; Sussman et al., 2003). They show that the KAB-based model performs better than the other state-of-the-art models in predicting the usefulness of answers (Liu X. et al., 2020).

Third, knowledge networks play an important role in knowledge sharing. Guan et al. (2017) argue that the knowledge elements' positions in the knowledge network can also influence a paper's citations. They find that a paper's centralities in both the collaboration network and knowledge network have an inverted U-shaped impact on citations. Further, they find that the structural holes of a paper in the knowledge network have a positive influence on its citations (Guan et al., 2017). Zhang et al. (2019) quantify the degree of knowledge recombination in the knowledge graph, and they also find an inverse U-shaped relationship between recombinant distance and a firm's innovation. Moreover, knowledge recombination gives rise to knowledge convergence. Liu N. et al. (2020) find that the intensity of science (i.e., the degree to which scientific knowledge is adopted in technology research and development) has an inverted U-shaped influence on knowledge convergence and the impact is moderated by relational embeddedness.

## 2.3 Knowledge Spanning and Knowledge Hierarchy

Questions' positions in the knowledge space shape their popular success (Askin and Mauskapf, 2017; Guan et al., 2017; Guan et al., 2018; Zhang et al. 2019). The features of popular questions tend to collectively signal its quality and reflect the taste of the public (Chen et al., 2014; Chua et al., 2015). The ensemble of different features can be represented with the knowledge space. Knowledge space is a multi-dimensional representation of various forms of knowledge (e.g., questions, patents, research articles). It constitutes a broader ecosystem of knowledge production and consumption. Each question's features determine its position and interrelations with the other questions in the knowledge space. The interrelations of questions help people organize and evaluate questions. To summarize, people perceive and evaluate questions according to their positions in the knowledge space.

By synthesizing the theories of knowledge combinations and the studies on category spanning (Hannan & Freeman, 1977; Hannan et al., 2007; Hannan, 2010; Hsu et al., 2009), we conceptualize the act of knowledge recombination on the Q&A website as *knowledge spanning*. Knowledge spanning is one particular kind of knowledge recombination. It reflects the distance of boundary spanning in the knowledge space. Category spanning is one of the most commonly adopted innovation strategies for recombination. Categorization plays an essential role in human behaviors. As we have illustrated, categorization is conducive to the retrieval, browsing, and creation of knowledge. As an indispensable part of the online knowledge market, social markers (i.e., folk taxonomy) are widely used to mark and classify questions. A category could be a specific field, thing, or concept. Questions are categorized and indexed by their categories of knowledge. The online knowledge market usually urges the user who asks questions to add categories to help questions reach those who are interested in these questions.

Categories help users organize and discover content in the knowledge market. Users can retrieve, filter, share, and organize questions through this flexible and unrestricted system. Just as Zuckerman has put it, "social objects are evaluated by legitimate categories" (Zuckerman, 1999, p.1398). The categories of questions have a large impact on the evaluation of the question quality. For example, Toba et al. (2014) find that training a type-based quality classifier is effective in discovering high quality answers. Hu et al. (2020) use topic models to extract keywords from articles and construct five keyword popularity features (i.e., topic popularity, published popularity, news popularity, web page popularity, video popularity) based on the information collected from ResearchGate, Google Scholar, and Google Trends. They show that these keyword popularity features can effectively improve the prediction models of highly-cited papers.

The social object of category spanning is penalized for "perceived quality and audience attention" (Keuschnigg & Wimmer, 2017). The objects related to multiple categories are difficult to interpret. Thus, the public may underestimate the quality of these social objects and pay less attention to them. However, the findings are not entirely consistent with each other. For example, Askin and Mauskapf (2017) find that the songs of optimal differentiation are more likely to succeed than the typical songs. Similarly, Shi et al. (2021) find that the tag distance has an inverted U-shaped influence on the popularity of questions.

There are many other bottlenecks of knowledge spanning. For example, Kuhn (1977) argues that education plays a crucial role in maintaining tradition and restricting innovation. It is difficult to transfer knowledge across groups and generations. Since knowledge accumulates over time, successive generations of innovators face an increasingly massive amount of accumulated knowledge. Human beings are the carrier of knowledge. The difficulty in the transferring of knowledge (i.e., human capital) becomes the bottleneck of innovations. Therefore, Jones (2008) argues that there is an increasing *burden of knowledge* for innovators. Gruber et al. (2012) find that a scientific education helps inventors carry out knowledge recombination across technological boundaries. Nevertheless, the breadth of inventors' knowledge recombination decays over time after their education.

Prior research on category spanning proposes two mechanisms to explain the negative consequence of category spanning (Keuschnigg & Wimmer, 2017). First, category spanning tends to reduce the niche fitness of cultural products. Category spanners are usually generalists who allocate their time and effort into multiple categories. Compared with the specialists who focus more on a particular field, they gain less experience, ability, and niche fitness. Second, category spanning tends to increase the audience's confusion towards cultural products. It is more difficult for the audience to form stable expectations for category spanners than non-spanners. There are cognitive difficulties when evaluating questions raised by knowledge-spanners. Since the questions of a high degree of knowledge recombination usually span multiple disciplines, people need to consume more cognitive resources to understand it. Keuschnigg and Wimmer (2017) find that the confusion mechanism is relatively more important than the fitness mechanism in explaining the spanning effect.

Going beyond the limit of recombination is as bad as falling short. In the perspective of Bourdieu's field theory of science, Foster et al. (2015) conceptualizes the choice of research question as strategic action and habitus. They find that conservative strategies are common, while innovative strategies are rare. Although innovative publications can achieve a higher impact than conservative ones, they face a more considerable risk of failing to publish. Specifically, they distinguished five scientific strategies: repeat bridges, repeat consolidations, new bridges, new consolidations, and jumps. More interestingly, they find that the impact of research depends quadratically on the proportion of each scientific strategy, whether innovative or not.

Similarly, Uzzi and Spiro (2015) find a reversed U-shape relation between the extent of recombination and musicals' success. Askin and Mauskapf (2017) argue that there is a tradeoff between similarity and differentiation. They measure the genre-weighted typicality of popular songs with Cosine similarity and find a reversed U-shape relation between typicality and popularity. Since the degree of spanning can be measured with 1-Cosine similarity and the reversed U-shape is symmetric, their findings suggest a reversed U-shape relation between boundary spanning and popularity. Using the number of answers to represent the popularity of a question, Shi et al. (2021) construct the knowledge tag network and examine the impact of distance on the popularity of questions on Stack Overflow. The tag distance measures how close the tags are in a question, that is the degree of knowledge spanning. Their findings suggest that the tag distance has an inverted U-shaped influence on the popularity of questions and this nonlinear influence is moderated by tag frequency (Shi et al., 2021). Taken together, knowledge spanning has a parabolic effect on the success of questions. Below a critical point, the increase of knowledge spanning would enhance the appeal of questions; otherwise, it would reduce the appeal of questions. Thus, we propose the following hypothesis:

*H1: there exists a reversed U-shape relation between the distance of knowledge spanning and the appeal of questions.*

Knowledge is also featured by its hierarchical structure (Cole, 1983). The similarity is just one dimension of knowledge spanning, and another dimension that is often overlooked is the hierarchy. Similar questions have a higher chance to be close to each other in the knowledge space. However, the attractiveness of questions is not only determined by the tradeoff between similarity and differentiation. Cole argues that knowledge is not uniformly distributed (Cole, 1983). Although it is a continuum, it can be divided into two main parts: the core of knowledge and the research frontier (Cole, 1983). The core usually consists of a small number of well-recognized theories. It is the starting point based on which people can produce new knowledge. The social process of evaluation connects the core and the frontier. As time goes by, only a tiny part of the frontier knowledge is still evaluated as valuable and eventually becomes the core of knowledge. There is greater consensus towards the core of knowledge in the scientific fields at the top of the hierarchy compared with that at the bottom. But there is no such difference for the consensus on the research frontier between the scientific field at the top and that at the bottom of the hierarchy (Cole, 1983).

The present research examines the hierarchical structure of knowledge. The representation of knowledge or thought is primarily organized in the form of language. Language is characterized by its hierarchical structure (Tang, X., et al., 2019). The American linguists Hayakawa and Hayakawa explain it with the ladder of abstraction (Hayakawa & Hayakawa, 1939). Human beings are capable of reasoning on different levels of abstraction. For example, the cow named Bessie living on a farm could be described on eight levels of abstraction (from concrete to abstract): process level (e.g., an object composed of atoms), name, Bessie, cow, livestock, farm assets, assets, wealth. Hayakawa et al. (1939) argue that the ladder of abstraction widely exists in our thought and communication. The kids of the farmer consider Bessie as a lovely cow. In comparison, the farmer describes Bessie as wealth. It is reasonable to infer that choosing different levels of abstraction will produce different social consequences.

Coarse-grained categories occupy higher positions in the hierarchical structure of knowledge. According to the level of abstraction, the categories used to label the types of questions on the Q&A website constitute a knowledge tree. Accordingly, questions can be classified into two types: abstract questions and concrete questions. Since abstract questions are framed in a more general way, they cover a much larger group of people. In contrast, the concrete questions can only cover a much smaller number of users. Yet, just as Hannan and Freeman (1977) argued, although the width of the niche market for concrete questions is narrower than that of abstract questions, they are more manageable, operational, and competitive. However, the superiority of the narrow niche is contingent upon the stability of the environment. If the environment is unstable, choosing a wide niche is a better strategy. Lerner and Lomi (2018) find that the Wikipedia articles with coarse-grained categories can get more attention but lower evaluations (e.g., they are not featured articles). Overall, the position of the question in the hierarchical structure, or the hierarchy of the knowledge, also plays an important role in knowledge sharing (Lerner & Lomi, 2018). Based on the analysis above, we propose the following hypothesis:

*H2: the extent of knowledge hierarchy influences the appeal of questions.*

The effect of boundary spanning is contingent on various social factors (Keuschnigg & Wimmer, 2017; Goldberg et al., 2016; Shi et al., 2021; Liu N. et al., 2020). Keuschnigg and Wimmer (2017) find that the impact of category spanning on the price of the film DVD is moderated by market context (e.g., the major films or not), product familiarity (e.g., past box-office success), and social-cultural distance (e.g., the regularity of co-occurrence). When these moderators are considered, the negative effects of category spanning are far from ubiquitous. The people who prefer typicality dislike the cultural products that span boundaries (Goldberg et al., 2016). Goldberg et al. (2016) construct a geometric model to measure users' taste for variety and typicality. Their findings suggest that the impact of cultural products' object atypicality (category spanning) on users' latent appeal is moderated by users' taste for variety and typicality.

The effect of category spanning is contingent on the *fuzziness* of categories (Kovács and Hannan, 2010). The strength of category boundaries reflects the *partiality of membership* in categorization research (Hannan, Pólos, and Carroll, 2007). If the partial assignments of membership become common, the boundaries will become blurred or fuzzy. Kovács and Hannan (2010) propose that when the category boundaries are fuzzy, spanning them will not increase the audiences' confusion. If so, the penalties of boundary spanning would be small or even negligible. Else, the penalties would be more significant. As a result, it is easier to span the categories with blurry boundaries. This idea is captured by the concept of *category contrast*. Low contrast or high fuzziness reduces the appeal of the object (Negro, Hannan, Rao, 2011). Since coarse-grained categories are in the higher positions of the hierarchical structure of knowledge, they are more abstract and do not have clear-cut boundaries. As a result, the effect of category spanning for the question with coarse-grained categories will be larger than that with fine-grained categories. Based on the logic above, we propose our hypothesis as follows:

*H3: the impact of knowledge spanning on the appeal of questions is moderated by the question's knowledge hierarchy.*

## 3. Methods

## 3.1 Data

We collect the data from Zhihu.com. Zhihu is the largest Q&A website in China as an alternative to Quora (Wikipedia, 2021a, 2021b). Quora, one of the most famous Q&A websites, has attracted more than 326 million registered users (Wikipedia, 2021a). Similarly, the number of registered users of Zhihu.com has reached 220 million by 2018 (Wikipedia, 2021b). In our dataset, there are 312,053 questions asked by the users on Zhihu.com from Dec 2010 to May 2018. Each question record contains the question id, the number of answers, the number of question followers, the content of the question, and a list of categories. Most questions (66%) have two or more categories.

We choose to study knowledge spanning on Zhihu.com, not only because it has a vast number of users in China and is playing an increasingly important role in the formation of public opinion, but also because Zhihu.com officially maintains a well-structured knowledge tree. The node of Zhihu.com's knowledge tree represents the category of questions, and the direct link from the father node to the child node suggests the hierarchical relationship between them. Based on the knowledge tree and the questions' category lists, we design our research as follows (see Figure 1):

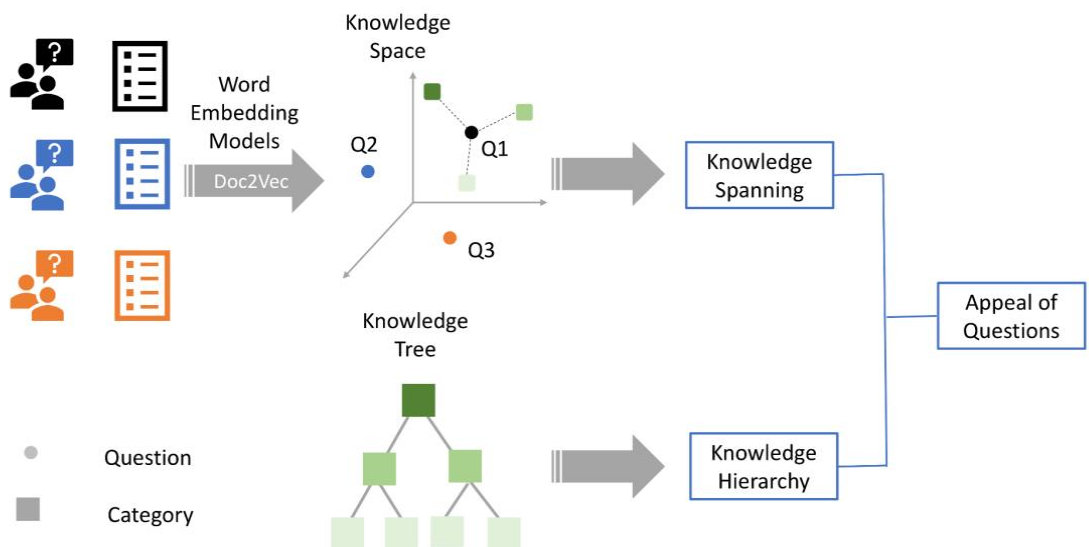

**Figure 1.** The Research Design of Knowledge Spanning

First, we construct the knowledge space using word embedding models. Viewing the list of categories appearing on the question webpage as a sentence, we employ the Word2Vec model to train a neural network model. After that, we can represent each category and question as an N-dimensional vector. Thus, we can measure the distance of knowledge spanning for each question based on its categories' positions in the N-dimensional knowledge space. Second, Zhihu.com officially maintains a knowledge tree for all the categories of the questions (N = 108,432). We can calculate the distance from each category to the root node in the knowledge tree and measure the knowledge hierarchy for each question. Third, we quantify the appeal of questions based on the number of followers of each question. Finally, we build up non-linear regression models to test our hypotheses.

## 3.2 Measures

***Knowledge Spanning***. Natural language processing researchers have made significant progress in representing relationships between words. The words in a corpus can be embedded as vectors into a dense, continuous, high-dimensional space. These vector space models, known collectively as *word embeddings models*, have attracted widespread interest among computer scientists and computational linguists due to their ability to capture and represent complex semantic relations. Word embedding models have inspired a wide range of item-context embedding models beyond words, ranging from images (Xian et al. 2016) and audio clips (Xie and Virtanen 2019) to graphs (Perozzi, Al-Rfou, and Skiena 2014; Grover and Leskovec 2016) and journals (Tshitoyan et al. 2019; Peng et al. 2020; Miao et al. 2021).

The ongoing development of neural networks and deep learning also provide potent tools (especially the word2vec algorithm) to explicitly model the latent knowledge space. Based on the knowledge space, the concept of knowledge spanning can be redefined. In our word embedding model, each category is represented as a vector in a vector space. The categories in similar contexts will be positioned nearby, whereas categories in distinct contexts will be positioned farther apart.

Figure 2 schematically illustrates the neural network structure of the word embeddings models. Word embedding models aim to represent all words from a corpus within the *k*-dimensional space that best preserves distances between $n$ words across $m$ semantic contexts. We set $k = 50$ in our study. As shown in Figure 2, the solution is a matrix with $n$ rows and $m$ columns. For a question with two categories ($T_a$ and $T_b$), we define the semantic similarity as the cosine similarity of their vectors in the knowledge space,

$$CosineSimilarity(T_a,\ T_b)\ = \frac{T_a \cdot T_b}{|T_a|\ |T_b|}.$$

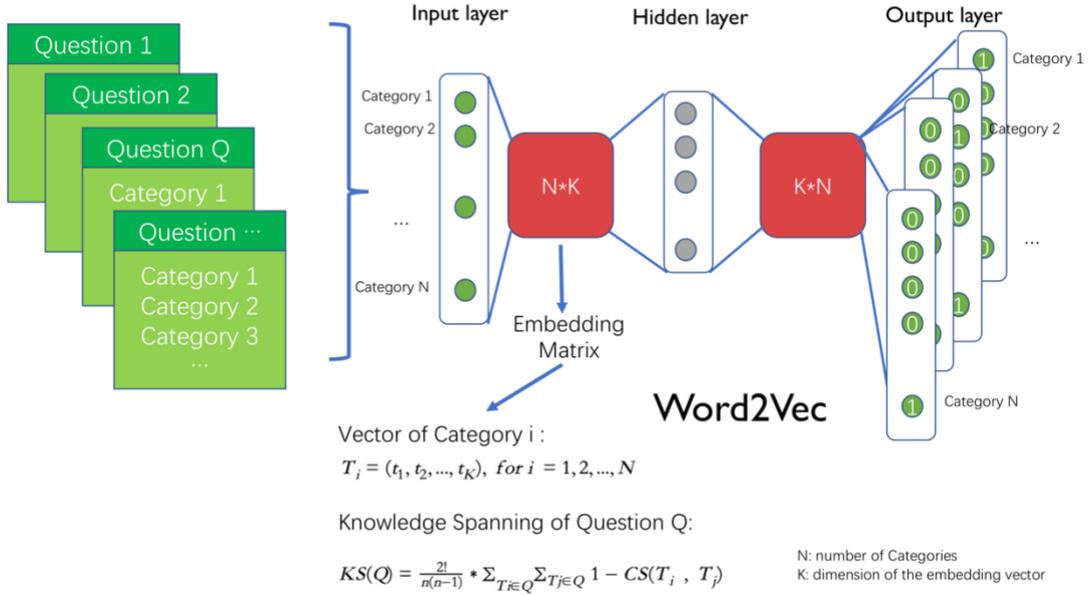

**Figure** 2. The Neural Network Model of Word Embeddings

Thus, the distance of knowledge spanning is $1\text{-}\ CosineSimilarity\ (W_a,\ W_b)$. For those questions with three or more categories, the distance of knowledge spanning in the knowledge space $KS(Q)$ is represented as follows:

$$KS(Q) = \frac{2!}{n(n-1)} * \Sigma_{Ti \in Q} \Sigma_{Tj \in Q} 1 - CS(T_i,\ T_j).$$

In the equation, $n$ is the number of categories for question $Q$. For questions with only one category, $KS(Q) = 0$. To Summarize, we can get the knowledge space with this method and measure the distance of knowledge spanning for each question. As the dimension of knowledge space is much larger than two, we need to conduct dimension reduction before visualization. Figure 3 shows the 2d knowledge space. Each node represents a category, and the nodes of the same category are visualized with the same color. Because the distribution of knowledge spanning is highly skewed, we transform it with its logarithmic form ($M$ = -2.06, $SD$ = 0.71).

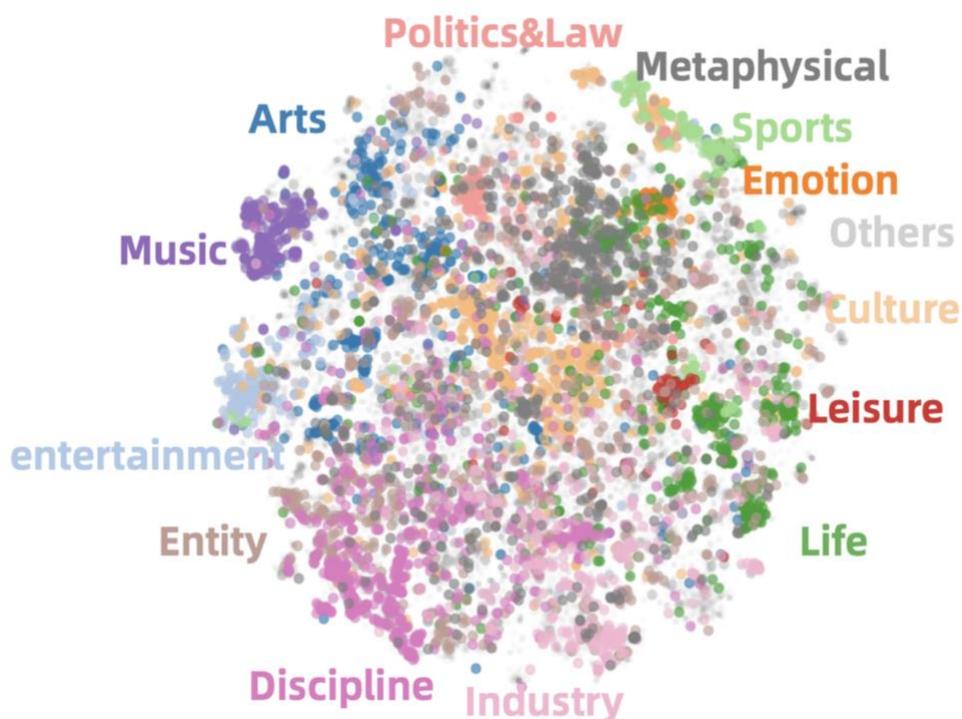

**Figure 3.** The Visualization of 2D Knowledge Space

**Knowledge Hierarchy**. Each category has a category page on Zhihu.com. The category page displays the number of followers, the number of discussions related to the category, a list of father categories, a list of child categories, a brief introduction of the category, a list of discussions, a list of highly-rated answers, and a list of new questions. Zhihu.com adopts a hybrid classification system to facilitate the categorization of questions. It was initially designed by the website developers and further improved by active users. First, when a user raises a question on Zhihu.com, she has to choose up to five categories for the question. Second, Zhihu.com will automatically suggest three categories based on the content of the question.

The knowledge tree of Zhihu.com is constructed based on the hierarchical relationship between different categories. In order to make the category structure concise, non-repetitive, and non-overlapping, Zhihu.com adopts the scientific taxonomy method to construct the knowledge tree. Based on the parent-child relationship, the categories form a directed acyclic graph (DAG). The root node of the knowledge tree is the top-level parent category of all other categories. It has six child nodes or sub-categories: subject, entity, metaphysics, industry, life/art/culture/activities, and unclassified.

We use the knowledge tree to measure the knowledge hierarchy of each question. The hierarchical level of categories reflects the granularity of questions. We set the hierarchical level of the root category as zero, the hierarchical level of its six sub-categories as one, and so on. The category on the higher hierarchical level in the knowledge tree has a lower abstraction level and the smaller granularity. The question's knowledge hierarchy (M = 4.71, SD = 2.41) is measured by the largest hierarchical level of its categories.

**Appeal of Questions**. Consistent with prior research, we measure the appeal of questions by content popularity. The popularity of specific content can be quantified by users' particular behaviors stimulated by the content. There are different ways to measure content popularity. For example, the popularity of a post can be expressed by the number of views, likes, followers, reposts, and replies (Ravi, S., et al., 2014; Toba, et al., 2014; Yao, et al., 2015). There is a strong correlation between these measurements (Wang & Wu, 2016; Wang et al., 2016). In this present research, we use the number of followers to measure the appeal of the question. However, similar to the other measurements of success, the distribution of the appeal of questions is also highly skewed (M = 833, SD = 5490). Thus, we transform it to the logarithmic form (M = 1.62, SD = 0.96). Table 1 illustrates the main variables and their statistical information in the research.

## 4. Results

The first hypothesis of this study proposes *a reversed U-shaped relation between the distance of knowledge spanning and the appeal of questions*. As shown in the first subplot of Figure 4, the appeal of questions has a reversed U-shaped relation with knowledge spanning. Using the bootstrap method, we get the average appeal of a question. Furthermore, the gray belt around the black line in Figure 4 covers 95% of the data. The curve increases when knowledge spanning (log) is less than -0.6, and it decreases when knowledge spanning (log) is larger than -0.6 (see Figure 4 all), which supports our hypothesis H1. We have also examined H1 across categories. The other subplots of Figure 4 demonstrate the relation between knowledge spanning and the appeal of questions for the fifteen most popular categories. The findings indicate that the inverted U-shaped relation exists across categories.

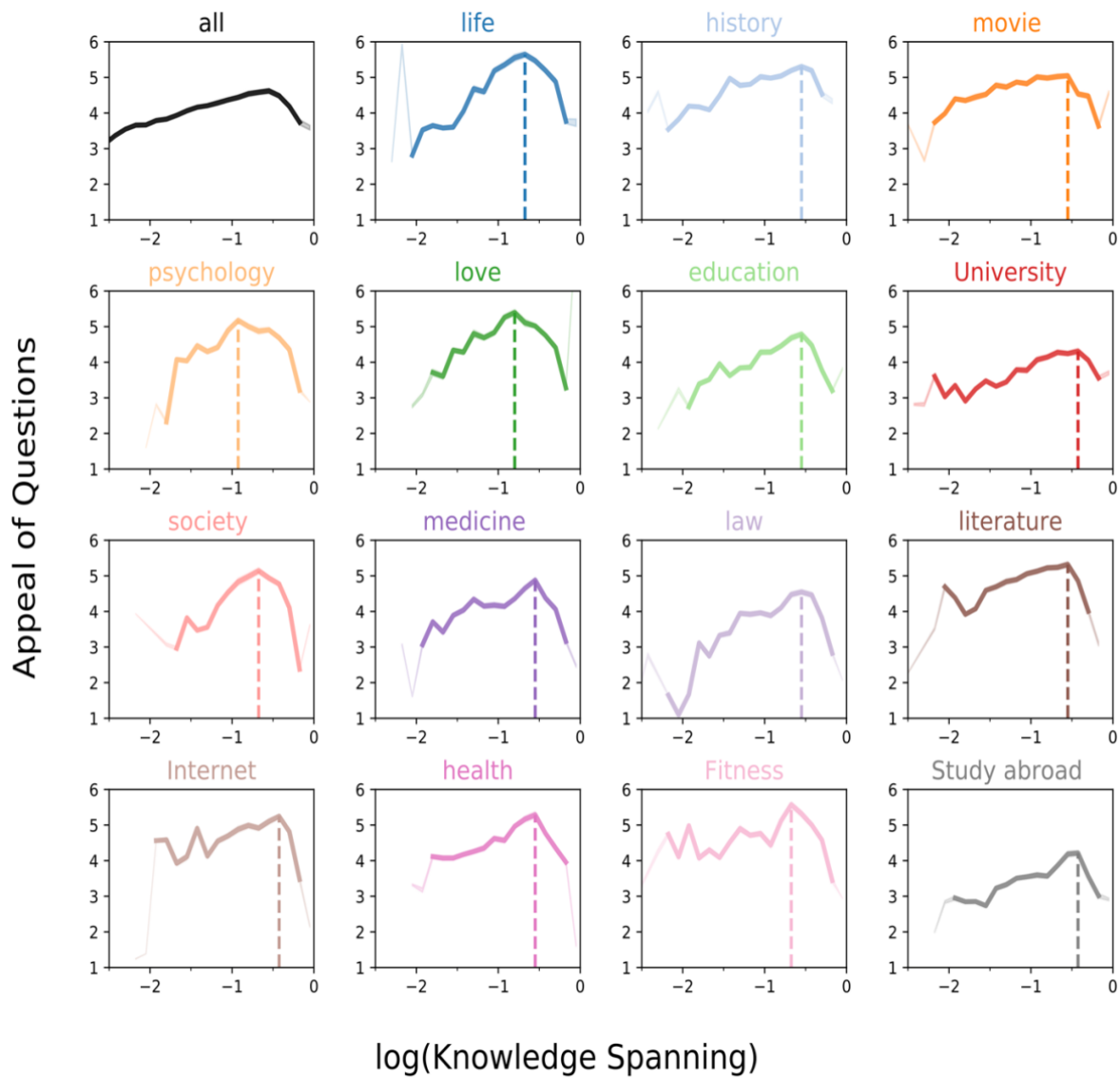

**Figure 4**. The Effect of Knowledge Spanning for the Questions of Different Categories.

**Table** 1. Multiple Linear Regression of the Appeal of Questions

| | Model 1 | Model 2 | Model 3 |
|---|---|---|---|
| Knowledge Spanning (log) | 417.131*** | 280.625*** | 334.198*** |
| | (2.123) | (1.831) | (5.091) |
| Knowledge Spanning (log)$^2$ | -68.074*** | -76.937*** | -42.513*** |
| | (2.123) | (1.771) | (6.281) |
| Hierarchy | | -0.010*** | -0.016*** |
| | | (0.002) | (0.002) |
| Knowledge Spanning (log) ×Hierarchy | | | -10.404*** |
| | | | (0.938) |
| Knowledge Spanning (log)$^2$ ×Hierarchy | | | -5.813*** |
| | | | (1.069) |
| Title Length | | -0.011*** | -0.011*** |
| | | (0.0003) | (0.0003) |
| Lasting Days | | 0.724*** | 0.724*** |
| | | (0.002) | (0.002) |
| Monday | | 0.007 | 0.007 |
| | | (0.009) | (0.009) |
| Tuesday | | 0.008 | 0.009 |
| | | (0.009) | (0.009) |
| Wednesday | | 0.019* | 0.020* |
| | | (0.009) | (0.009) |
| Thursday | | -0.0003 | -0.0001 |
| | | (0.009) | (0.009) |
| Friday | | 0.001 | 0.001 |
| | | (0.009) | (0.009) |
| Constant | 4.018*** | 0.359*** | 0.398*** |
| | (0.003) | (0.015) | (0.015) |
| Adjusted R$^2$ | 0.089 | 0.389 | 0.390 |

*Note:* * $p < 0.05$, ** $p < 0.01$, and ***$p < 0.001$. The values in parentheses denote standard errors.

Further, we build multiple linear regression models to formally test our hypotheses (see Table 1). First, we include the knowledge spanning (log) and its square form into our model to test the parabolic relation (see Model 1 in Table 1). Second, we include knowledge hierarchy, the length of question title, lasting days (log), and the days of the week as control variables (see Model 2 in Table 1). Third, we include the nonlinear interaction terms between knowledge hierarchy and knowledge spanning (see Model 3 in Table 1). According to Table 1, the coefficients of the square of knowledge spanning (log) are negative, indicating a U-shaped relation. Thus, hypothesis H1 is well supported.

The second hypothesis argues that there is an association between knowledge hierarchy and the appeal of the questions. According to Model 2 in Table 1, the main effect of knowledge hierarchy on the appeal of questions is negative (B = -0.01, p < 0.001). In other words, the more concrete the question, the less appealing it is. Thus, H2 is well supported.

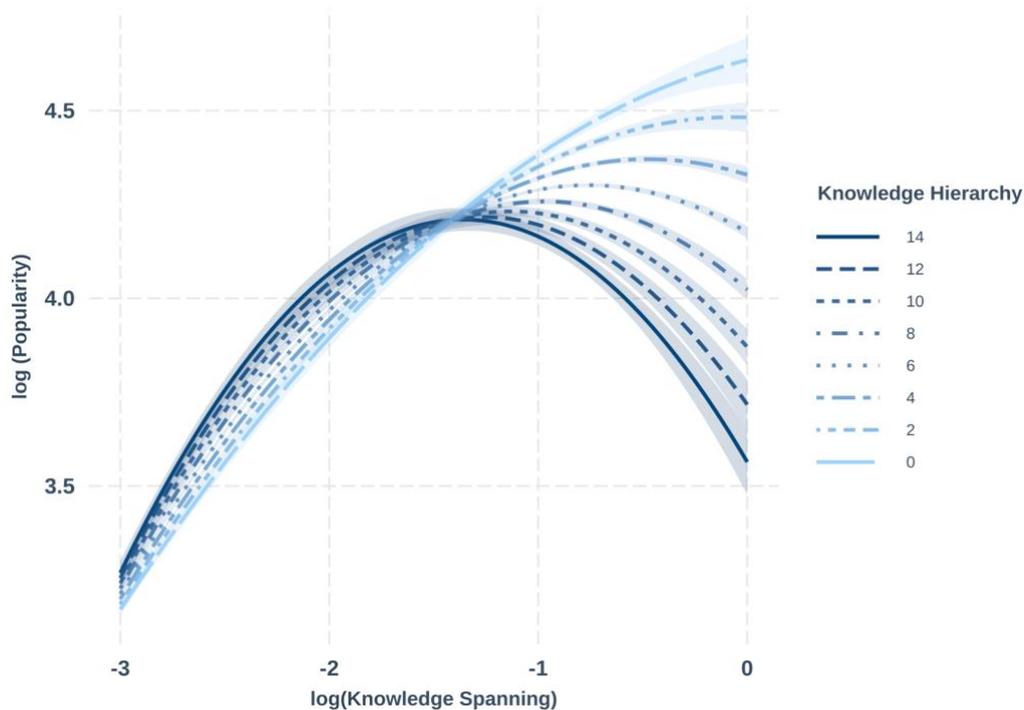

**Figure 5**. Regression of Question Appeals on Knowledge Spanning

The most important hypothesis of this study (H3) proposes that the nonlinear effect of knowledge spanning is moderated by the knowledge hierarchy of questions. As model 3 in Table 1 shows, there are significant interactions between knowledge spanning and question's knowledge hierarchy. To better understand the moderation effect, we visually illustrate it in Figure 5. Apparently, the existence of the inverted U-shape depends on the knowledge hierarchy of questions. If the knowledge hierarchy of questions is larger than 6, the inverted U-shape gets more prominent. However, when the knowledge hierarchy of questions is smaller than 6, the inverted U-shape relationship disappears quickly. Thus, H3 is well supported.

## 5. Discussions and Conclusions

Drawing on the theories of knowledge recombination, we reformulate the hypothesis of knowledge spanning in the context of the Q&A website. Using the data collected from a large Q&A website, we examine how the variation of knowledge spanning affects the questions' success in attracting public attention. Our findings indicate that good questions have the capability to achieve the tradeoff between similarity and differentiation. The optimal differentiation in the knowledge space shapes the success of questions on the Q&A website (Askin and Mauskapf, 2017; Shi et al., 2021). However, the nonlinear influence of knowledge spanning is contingent upon the position of questions on the ladder of abstraction.

### 5.1 Theoretical Contributions

This study makes an essential contribution to the theories of knowledge recombination by supplying a new perspective on the popular success of questions in the online knowledge market. We confirm the nonlinear knowledge spanning effect. Admittedly, knowledge recombination plays a crucial role in innovation (Schumpeter, 1934; Nelson & Winter, 1982; Koesterler, 1964; Zhang et al., 2019; Shi et al., 2021). There is no making without breaking. This phrase nicely expresses Schumpeter's notion of *creative destruction* (Schumpeter, 1942). If a question spans a long distance in the knowledge space, it would be unvalued or even completely ignored. In contrast, if a question only spans a very short distance in the knowledge space, it may suffer from a lack of novelty. However, there is a tradeoff or essential tension between tradition and innovation. The tradeoff is achieved by negative feedback.

The marginal return of information spanning diminishes with the increase of knowledge spanning. Moreover, when information spanning is above a threshold, the marginal return even becomes negative. Just as our research has demonstrated, excessive knowledge spanning will hinder the appeal of questions. If the risk of recombination is one kind of destruction of tradition, the negative feedback suppresses such destruction and enhances the stability of the knowledge market.

Interestingly, the existence of the inverted U-shaped effect is contingent on the knowledge hierarchy of questions. According to our results, only when the knowledge hierarchy of questions is very large (i.e., concrete questions) can a noticeable inverted U-shaped effect be observed. As the knowledge hierarchy of questions decreases, the negative feedback is significantly weakened. When the knowledge hierarchy of questions is smaller than four, the negative feedback almost disappears. Note that the mean value of knowledge hierarchy is relatively small ($M = 4.71$) on the Q&A market. In other words, people would like to ask relatively more abstract questions. Therefore, the negative feedback is minimal on the Q&A website. Consequently, the Q&A website is more tolerant or supportive of knowledge spanning.

Compared with the situation of high knowledge spanning, the moderation effect of knowledge hierarchy is much smaller when knowledge spanning is below a threshold. The questions of low knowledge hierarchy lost their advantage over those questions of high hierarchy. In other words, when knowledge spanning is below a threshold, people prefer concrete questions to abstract ones. We conjecture that the essence of the problem lies in the contest between different mechanisms driving the moderation effect. On the one hand, since a low knowledge hierarchy lacks contrast, it would not hurt the clarity of questions. As a result, the penalties associated with low knowledge hierarchy are smaller than that of high hierarchy. Therefore, the low hierarchy promotes the effect of knowledge spanning. This mechanism can be conceptualized as the *contrast hypothesis* (Kovács & Hannan, 2010).

On the other hand, the low knowledge hierarchy implies that the audience members attracted by the same label are heterogeneous. In other words, they share fewer common features and are more likely to disagree with each other. In this case, the low knowledge hierarchy restrains the effect of knowledge spanning. These mechanisms can be conceptualized as the *heterogeneity hypothesis* (Negro,

Hannan, and Rao, 2011). When knowledge spanning is below a threshold, the promoting effect is smaller than the inhibiting effect, and the questions of low knowledge hierarchy lose their competitive advantage. When knowledge spanning is above a threshold, the promoting effect is much more powerful, and the questions of low knowledge hierarchy are more successful than those of high knowledge hierarchy. To sum up, concrete questions are more prevalent when knowledge spanning is small, while abstract questions are much more popular when knowledge spanning is large.

## 5.2 Practical Implications

First, human beings acquire knowledge through question and answer, but the popular success of questions is socially constructed (Barabási, 2018). The feedback mechanism is strongly driven by the social process of evaluation (Cole, 1983). According to the level of knowledge spanning, the innovation search strategies can be classified into two categories: the exploitation strategy (small recombination) and the explorative strategy (large recombination). Surprisingly, the stock market prefers unfamiliar explorative patents to incremental exploitative patents (Fitzgerald et al., 2021). As a result, although the firms focusing on the exploitation strategy generate much better operating performance in the short term, they are undervalued by the stock market. We argue that the cognitive bias towards knowledge spanning has a similar effect on the Q&A website. Social knowledge as public goods is similar to fictitious capital (e.g., stocks). As a result, the Q&A website has much fewer constraints on knowledge recombination.

Second, and more importantly, maintaining a hierarchy tree of categories is a good idea for the knowledge market. The market prefers high-contrast inventions (Kovács et al., 2021). Potential users' preference of knowledge not only depends on its value, but also on the degree of contrast. For example, Similarly, Kovács et al. (2021) find that the patents of the high-contrast categories can attract more attention and citations compared with those of the low-contrast categories (Kovács et al., 2021). Consistent with Kovács et al. (2021), our findings also indicate that the knowledge hierarchy can flatten or even reverse the U-shaped impact of knowledge spanning. Thus, the knowledge market can better encourage knowledge spanning by controlling the knowledge hierarchy.

Third, knowledge recombination occurs in social fields (Bourdieu, 1975). The social field of the Q&A website can be described with the knowledge space. And the knowledge space helps people navigate in it. With the aid of the knowledge space, we can locate the position of each category, question, and person. However, there is a challenge of capturing agents' positions in the social fields. Prior research of categorizations attempts to solve this question by constructing the feature space and label space (Pontikes & Hannan, 2014). When the number of variables is large, there exists a problem of representation. For example, there are hundreds of thousands of question categories ($N = 108,432$) on the Q&A website. If we represent them with dummy coding or one-hot encoding, most values of these variables are zero. As a result, it is impossible to compute the distance between two questions. Fortunately, the perspectives of geometry and network provide a powerful lens to investigate knowledge recombination (Uzzi & Spiro, 2015; Foster et al., 2015; Kozlowski et al., 2019; Min et al., 2021; Li, et al. 2020; Lee, et al., 2021). By building a geometric model of knowledge spanning, we embed questions into a high dimensional geometric space, construct the knowledge space, and quantify the distance of knowledge spanning on the Q&A website.

## 5.3 Limitations

We acknowledge the limitations of this research which pave the road for future research. First, we have only examined knowledge spanning in asking questions on the Q&A website. How to understand the knowledge spanning in answering questions on the Q&A website remains a question. Future research should pay more attention to answer this question. Second, we quantify similarity-based boundary spanning with word embedding models and capture the hierarchy-based boundary spanning with the network analysis based on the knowledge tree. Is it possible to unit hierarchy and similarity within one model? One solution is to embed the categories of questions into a hyperbolic space rather than a Euclidean space. Thus, we can observe both hierarchy and similarity within the same space (Papadopoulos et al., 2012; Li et al., 2020). Third, although we believe our findings could be generalized to the other types of the information market, we limit our analysis to the Q&A website. It is necessary to test the findings of this study in different forms of knowledge.

## 5.4 Conclusions

In all, this research confirms that knowledge spanning has a parabolic effect on the success of questions. However, the parabolic impact of knowledge spanning is contingent on the moderation of knowledge hierarchy. The social evaluation process plays a crucial role in linking the knowledge in communication and the knowledge core (Cole, 1983). Human beings evaluate questions from two dimensions: one is the similarity between questions, the other is the hierarchy of the questions on the ladder of abstraction. Our study demonstrates that the Q&A website and its users are more supportive of knowledge recombination. The advantage of the knowledge market lies in constructing a knowledge space and building a hierarchical knowledge tree. Moving up the ladder of abstraction helps break the boundaries of knowledge space.